\documentclass[aps,prl,twocolumn,groupedaddress,showpacs]{revtex4}
\usepackage{epsfig}

\def\p{{\bf p}}

\def\8{\infty}
\def\oh{\frac{1}{2}}

\def\undertext#1{\vtop{\hbox{#1}\kern 1pt \hrule}}

\def\VEa#1{\left\langle\,#1\,\right\rangle}

\def\pp#1{\frac{\partial}{\partial#1}}

\def\p2byp#1#2{\frac{\partial^2#1}{\partial#2^2}}

\def\be{\begin{equation}}
\def\ee{\end{equation}}
\def\bea{\begin{eqnarray} & &}
\def\eea{\end{eqnarray}}

\def\rf#1{(\ref{#1})}

\def\rf#1{(\ref{#1})}

\def\rfs#1{Eq.~\rf{#1}}

\begin{document}


\title{Excitations of the One Dimensional Bose-Einstein Condensates in a Random Potential}


\author{V. Gurarie$^1$, G. Refael$^2$, J. T. Chalker$^3$}

\affiliation{$^1$Department of Physics, University of Colorado,
Boulder CO 80309, USA}
\affiliation{$^2$Department of Physics, California Institute of Technology, Pasadena CA 91125, USA}
\affiliation{$^3$Rudolf Peiers Centre for Theoretical Physics, Oxford University,
1 Keble Road, Oxford OX1 3NP, UK}

\date{\today}

\begin{abstract}
We examine  bosons hopping on a one-dimensional lattice in the presence of a random potential at zero temperature. Bogoliubov 
excitations
of the Bose-Einstein condensate formed under such conditions are localized, with the localization length
diverging at low frequency as $\ell(\omega)\sim 1/\omega^\alpha$. We show that the well known result $\alpha=2$ applies
only for sufficiently weak random potential. As the random potential is increased beyond a certain strength, $\alpha$ starts decreasing. 
At a critical strength of the potential, when the system of bosons is at the transition from a superfluid to an insulator, $\alpha=1$. 
This result is relevant for understanding the behavior of the atomic Bose-Einstein condensates in the presence of random
potential, and of the disordered Josephson junction arrays. 
\end{abstract}

\maketitle


One of the most challenging problems of quantum many body physics is
the behavior of stongly interacting matter in a disordered
environment. In this paper we investigate the universal
properties of superfluids in such systems, near the superfluid
insulator transition. Interest in
this problem arises in many independent contexts, in work on granular superconducting films and wires
\cite{GoldmanHaviland,DynesG,Lau}, Helium condensates in vycor \cite{Reppy2},
and recent experiments on Bose condensates in optical traps. In
particular, issues such as the expansion of a noninteracting Bose
condensate through a random potential~\cite{Aspect2006}, excitations
in an interacting Bose Einstein condensate in a random
potential~\cite{Pavloff2006,Aspect2007}, and the possibility of the
observation of the Bose glass phase~\cite{Inguscio2007,Inguscio2007a}
were explored in very recent theoretical and experimental papers. As important is the
possibility of investigating the behavior of disordered superconductors in a
controlled fashion using Josephson junction arrays, as in
Refs. \cite{Delsing1998,Lehnert2007,Clarke}. In low dimensional quantum
systems, where symmetry broken phases are very fragile, we expect the most dramatic manifestations of the interplay
of disorder and interactions. The existence of the Bose-glass phase
was established in Refs. \cite{Giamarchi1988,Weichman1989},
where the scaling and renormalization group (RG) picture of the
1d superfluid-insulator transition at weak disorder was also
established. Recently, much theoretical progress was afforded through
real-space RG approaches in the case of dissipative \cite{Vojta2007}
and closed \cite{Altman2004,Altman2007} bosonic chains, where the properties of
the SF-insulator transition at strong disorder were established.

In this paper, we study the excitations of the superfluid phase in a
bosonic chain with a strongly random potential and interactions, near
the SF-insulator transition. Capitalizing on the real-space RG
understanding of this transition \cite{Altman2004,Altman2007}, we analyze the
{ localization length of phonons} (i.e., Bogoliubov quasiparticles) as a
function of their frequency and wave number. Deep in the superfluid
phase, when the random potential is weak, the phonon localization
length $\ell(\omega)$ at small $\omega$ diverges as~\cite{Ishii1973,Ziman1982,Pavloff2006}
\begin{eqnarray} \label{eq:ldef} \ell(\omega) &\sim& 1/{\omega^\alpha} ,\\
\alpha & = & 2.  \label{eq:locstand}
\end{eqnarray}

This result, in particular, formed the basis of the analysis in Refs.~\cite{Pavloff2006,Aspect2007}.
Using the renormalization group analysis of
Ref.~\cite{Altman2004} and the study of random elastic chains of
Ref.~\cite{Ziman1982}, we show that 
\rfs{eq:locstand} 
does not apply
everywhere in the superfluid regime. In a finite region of parameter
space on the superfluid side near the superfluid-insulator transition,  
\rfs{eq:locstand} 
fails, and is replaced by the law 
\be  \label{eq:locus}
\alpha=g,
\ee
where $1 \le g \le 2$. The meaning of the parameter $g$ will be elucidated
later in the paper. Furthermore, as the system approaches the transition to
the insulating regime, $g$ decreases. Exactly at the transition
$g=1$, and  Eq. (\ref{eq:ldef}) acquires a correction to scaling:
\be \label{eq:loctran}
\ell(\omega) \sim \left( {\ln^2 \omega}\right)/{\omega}.
\ee
Eqs.~\rf{eq:locus} and \rf{eq:loctran} constitute the main result of
our paper. 

Our analysis begins by considering a one-dimensional
disordered Bose-Hubbard model with many particles per site. Its Hamiltonian is
\be \label{eq:ham}
\hspace{ -.4cm} H = \sum_k  \left[ \frac{U_k}2 \left(-i \pp{\phi_k} + n_k \right)^2- J_k \cos \left( \phi_{k+1}-\phi_k \right) \right].
\ee
This Hamiltonian describes a chain of sites, connected to their
nearest neighbors by a Josephson hopping with a random strength,
$J_k$. $U_k$ is the strength of the onsite repulsion, and
$n_k \in \left[-\oh, \oh \right]$ are random offset charges. The hopping, charging and offsets are
randomly distributed with probability densities $P_J(J)$, $P_U(U)$, and $P_n(n)$.

In the strong-disorder limit a real space
renormalization group analysis can be employed to gradually eliminate sites with anomalously large
$J_k$ or  or local charging gap, $\Delta_k=U_k(1-2|n_k|)$
\cite{Altman2004,Altman2007}. The remaining sites are described by the same
Hamiltonian \label{eq1} but with the renormalized probability
distributions. The system of \rfs{eq:ham} then emerges as either
a superfluid or an insulator; the latter could be either a Mott
insulator, a Mott glass, Bose glass, or random-singlet glass, depending on the strength, relative
and absolute, of various types of disorder present. If the bosonic system is a superfluid, the distribution of $J$
renormalizes towards the universal limiting function
\be \label{eq:PJ} 
P_J(J)=C J^{g-1}. 
\ee
with $C$ providing normalization. The superfluid is described by $g\ge
1$ with its value decreasing as the critical point at $g=1$ is
approached; in particular, as disorder increases, g decreases. At the same time,
 \be 
P_U(U) \sim \frac{1}{U^2} \exp \left( - \frac{\Omega f}{U} \right),
\ee
where $f$ flows to $0$ along the renormalization group trajectories
and $\Omega$ is the decreasing UV cutoff scale of the
renormalized Hamiltonian, i.e., its largest hopping or gap.
We now proceed to show that the same parameter $g$ appearing in the
distribution (\ref{eq:PJ}) controls the localization length of low-frequency phonons, as
expressed in Eq. (\ref{eq:locus}).

At the final stages of the renormalization, as long as $g \ge 1$, 
 the system is a superfluid and the possibility that the phase difference at adjacent sites slips through $2\pi$ can be safely ignored. Moreover, since the Hamiltonian is no longer periodic
in $\phi_k$ it is now possible to do a gauge
transformation to eliminate $n_k$ in \rfs{eq:ham}, marking the ability
of the superfluid to screen arbitrary offset charges. 
Then one
may expand the cosine, to find the effective
Hamiltonian (up to an unimportant additive overall constant)
\be \label{eq:hamred}
H = \sum_k  \left[ - \frac{U_k} 2 \frac{\partial^2}{\partial \phi_k^2} + \frac{J_k}{2} \left(\phi_{k+1}-\phi_k \right)^2
 \right].
\ee
The Hamiltonian Eq.~\rf{eq:hamred} is quadratic, and thus we obtain full information
by analyzing it at the classical level. Its classical equations of motion are
\be \label{eq:eqm} 
\omega^2 U_k^{-1} \phi_k  =J_k \left(\phi_{k}-\phi_{k+1} \right) + J_{k-1} \left( \phi_{k}-\phi_{k-1} \right),
\ee  
where $\omega$ is the angular frequency. 
These describe  phonons in a chain of random masses connected by random springs. The masses
are $ m_k \sim  1/U_k$, while the spring constants are proportional to $J_k$. 
Ref.~\cite{Ziman1982} presented the solution to this problem (referred to as
``Dyson type II") for the case when
$P_U(U)=\delta(U-U_0)$ (that is, nonrandom uniform charging energies $U_k=U_0$) and $P_J(J) = CJ^{g-1}$ with $J \in [0, J_0]$.  It was found that the average density of states is constant at low frequency,
\be \label{eq:DoS} 
\rho(\omega) =\VEa { \sum_n \delta \left(\omega-\omega_n \right) } \sim \ {\rm const}.
\ee
Here $\omega_n$ are the frequencies of the phonons described by
\rf{eq:eqm} and brackets denote averaging over random $J_k$. A  constant density of states at low frequency is of course a feature
shared with non-random elastic chains. The phonons which are
the solutions of \rfs{eq:eqm}, however, are all localized. Their localization
length obeys \rfs{eq:ldef} (at small $\omega$) with
\be \label{eq:Ziman}
\begin{array}{cc}
\alpha=2, \, g \ge 2; & \alpha = g, \, 1 \le g \le 2.
\end{array}
\ee
These results are compatible with our claim in beginning of this paper, Eqs.~\rf{eq:locus} and
\rf{eq:loctran}. Nevertheless, the uniform-$U$ treatment leading to
Eqs. (\ref{eq:Ziman}) cannot be considered a derivation of
the localization length results in our problem: the random boson
problem has charging energies $U_k$ which are also
randomly distributed. Below we show that the results given in
Eqs.~\rf{eq:DoS} and \rf{eq:Ziman} are valid even if $U_k$ are random,
as long as the probability of observing anomalously small $U_k$ is not
too large. In addition we show that the uniform-$U$ results for $g=1$
exhibit strong corrections to scaling [see Eq. (\ref{loctran})].


Let us first confirm that the fully random Bosonic chain, Eq. (\ref{eq:hamred}), also
has a finite constant density of states at low frequencies, as in
\rf{eq:DoS}. Consider the classical
ground state of a system of $N+1$ sites described by
\rfs{eq:eqm}, where the first grain's phase $\phi_0$ is kept fixed at
$\phi_0=0$, and a force $h$ is applied conjugate to the phase $\phi_N$ of the last grain. The equilibrium values of the variables $\phi_k$ can be found
by minimizing the energy
\be \label{eq:ener}
E= \oh \sum_{k=0}^{N-1} J_k \left( \phi_k -  \phi_{k+1} \right)^2 - h  \, \phi_N.
\ee 
which yields:
\be \label{eq:res}
\phi_k = h \sum_{l=0}^{k-1} J^{-1}_l, \, 0 <  k \le N.
\ee
Alternatively, $\phi_k$ can be computed in the following way. Introducing the variables $\psi_k =\phi_k/\sqrt{U_k}$, we can find $\psi_N$ by inverting the matrix ${\cal H}_{kl}$ defined by the expression
\be
 \oh \sum_{k=0}^{N-1} J_k \left( \psi_k \sqrt{U_k} -\psi_{k+1} \sqrt{U_{k+1}} \right)^2  \equiv
 \oh\sum_{k,l=1}^{N} {\cal H}_{kl} \psi_k \psi_l. 
\ee
Then 
\be \label{eq:res2}
\phi_k = \sqrt{U_k U_N} \, G_{kN}~ h.
\ee
where $G$ is a matrix inverse to ${\cal H}$.  In particular, we are interested in $k=N$ case when \rfs{eq:res2} can be rewritten as
\be \label{eq:res1}
\phi_N = h  \sum_n \frac{{ C_n\left[\phi_N^{(n)} \right]}^2}{\omega_n^2}, \
C_n = \left[ {\sum_{k=1}^{N} \frac{ \left( \phi^{(n)}_k \right)^2 }{U_k}} \right]^{-1}.
\ee
Here $\phi_k^{(n)}$ are the normalized solutions to the eigenmode equation
\rfs{eq:eqm} with the boundary conditions $\phi_0=0$ at the beginning
of the chain, and with the frequency $\omega_n$, labelled by the index
$n$. 

Next we compare the two expressions  for $\phi_k$ at $k=N$, \rfs{eq:res}  and \rfs{eq:res1}. We observe that for the probability distribution $P(J) = J^{g-1}$, as long as $g >1$, \be \label{eq:lin} \left<  \phi_N \right> = h 
\left<  {\sum_{k=0}^{N-1} J_k^{-1}} \right>  \sim N. \ee

On other other hand, 
\be \label{eq:em} \phi_N = N \int d\omega\, \rho(\omega)\, \frac{{ C_\omega \left[\phi_N^{(\omega)} \right]}^2}{\omega^2},
\ee
where $\phi^{(\omega)}$ refers to the eigenmode at frequency $\omega$,
and $\rho(\omega)$ is the density of states. Clearly, unless the
density of states is strongly suppressed at small $\omega$, the
integral \rf{eq:em} diverges due to small $\omega$ contributions. At
small $\omega$, the localization length exceeds the system size, thus $\phi_N^{(\omega)} \sim 1/\sqrt{N}$. 
At the same time, $\left< 1/U_k \right>$ is finite, which means that $C_\omega$ is both $\omega$ and $N$ independent. 
Suppose $\rho(\omega) \sim \omega^\gamma$, where $\gamma<1$. Then
\be \phi_N \sim \int d\omega\, \frac{\rho(\omega)}{\omega^2} \sim \frac{\rho(\omega_0)}{\omega_0},
\ee
where $\omega_0$ is the smallest frequency of the system, which can be found by
\be 
\int_0^{\omega_0} d\omega \, \rho(\omega) \sim \frac{1}{N}, \  \omega_0 \sim \frac 1 {N^{\frac{1}{1+\gamma}}}.
\ee
This in turn gives
\be
\phi_N \sim N^{\frac{1-\gamma}{1+\gamma}}.
\ee
Comparison with \rf{eq:lin} reveals that $\gamma=0$, i.e., \rfs{eq:DoS}.


We now return to the localization length. First, consider
the case of weak random $J_k = J_0 + \delta J_k$ and $U_k=U_0+\delta
U_k$. Treating $\delta J_k$ as a perturbation, it is easy to find the
localization length following
Refs.~\cite{John1982,Gurarie2003a}. Indeed, the mean free time can be
found by the Fermi golden rule, to go as $\tau^{-1} \sim \omega^2$,
while the mean free path goes as $\sqrt{J_0 U_0} \tau$. The
localization length is proportional to the scattering length in 1D,
thus $\ell(\omega) \sim \omega^{-2}$.

This calculation however ignores the possibility of the wave scattering off the anomalously small $J_k$ or $U_k$. Indeed, suppose we have a ``weak link" in \rfs{eq:eqm} where $J_{\rm weak \ link}=j$ on that link is much smaller than $J$ on other links, $j \ll J$. It is straightforward to check that the phonons with wave vector $q \gg j/J$ get reflected off this weak link, while those with wave vector $q \ll j/J$ pass straight  through.
This is easiest to see if  we solve \rfs{eq:eqm} with the assumptions that all $J_k$ are equal to $J_0$, while that of the weak link is $j \ll J_0$ and all the $U_k$ are equal to $U_0$. Then the transmission coefficient through the weak link is given by
\be \label{eq:tran} T = \frac{1}{1+q^2 \frac{J_0^2}{4 j^2}},\ee
where $q$ is the dimensionless wave vector which is assumed to be
small, or $|q| \ll \pi$. $T$ tends to $1$ at small $q$, and to $0$ at large
$q \gg j/J$.

It thus follows that a phonon with wave vector $q$ cannot have
a localization length bigger than the average distance between the
``weak links" with the strength of their couplings no bigger than
$q$ divided by the density of states. Using \rfs{eq:PJ} we can estimate the average separation between such weak links. We find
\be \label{eq:vaska}
\int_0^j dJ ~J^{g-1} \sim \frac{1}{\ell},
\ee
where $\ell$ is the average distance between the weak links $j$. 
This gives $\ell \sim 1/j^{g}$. Since $j \sim q$, and $q \sim \omega$
due to \rfs{eq:DoS}, the localization length is bounded from above by
$\ell\sim 1/\omega^g$, thus we arrive at our result, \rfs{eq:locus}. 

Scattering off the small $U_k$ can also reflect the short wavelength
waves.  Taking all $U_k$ equal to $U_0$, while the ``heavy link"
(recall that the ``masses" are inversely proportional to $U_k$) $U_k$
equal to $u \ll U_0$, and taking all the $J_k = J_0$ gives the
transmission coefficient
\be
T=\frac{1}{1+\frac{U_0^2}{4 u^2} q^2},
\ee
equivalent to \rf{eq:tran}. 
This, however, does not lead to any corrections to \rfs{eq:locus}. Indeed, using the same arguments as preceeding \rfs{eq:vaska}, we find 
\be
\int_0^u \frac{dU}{U^2} \exp \left( - \frac{\Omega f}{U} \right) \sim \exp \left( - \frac{\Omega f}{u} \right) \sim \frac{1}{\ell}. 
\ee
Here $\ell$ is the typical distance between these ``heavy" links. 
Again taking $u \sim q \sim \omega$, we find
\be
\ell \sim \exp \left( \frac{\Omega f}{\omega} \right).
\ee
This estimate is much bigger than \rfs{eq:locus} and thus the real localization length \rfs{eq:locus} remains unaffected.
This concludes the derivation of Eqs.~\rf{eq:DoS} and Eqs.~(\ref{eq:locstand},\ref{eq:locus},\ref{eq:loctran}).

The analysis of $\ell(\omega)$ above assumed that we probe the phonon
modes of the superfluid in the very end of the renormalization group
flow, once the power law that controls the distribution $P_J(J)\sim
J^{g-1}$, has already attained its fixed-line
value. While this is valid in the limit of $\omega\rightarrow 0$,
corrections to scaling may arise near the critical
point. Refs. \cite{Altman2004, Altman2007} allows us to consider the
corrections to this analysis arising from the flow to the SF fixed
line. The RG flow for the generic-disorder case is given by:
\be
\frac{df}{d\Gamma}=f(1-g), \ \frac{d g}{d\Gamma}=-\frac{1}{2}f g,
\ee
where $\Gamma$ is the logartihmic RG flow parameter. In the region close to the critical point,
$f=0,\,g=1$, we can solve these equations approximately to
give: 
\be
\begin{array}{cc}
g \approx 1+\epsilon+ \frac{2 \epsilon }{ e^{\epsilon \Gamma} -1}, & f\approx
\epsilon^2 \frac{4  e^{\epsilon \Gamma}}{\left(e^{\epsilon \Gamma} -1 \right)^2} 
\end{array}
\label{off-crit}
\ee
for disorder realizations that flow to $g=1+\epsilon$ with $\epsilon \ll 1$. Flows
that terminate at the critical point, however, are given approximately
by:
\be
\begin{array}{cc}
g \approx 1+\frac{2}{\Gamma}, & f\approx \frac{4}{\Gamma^2}.
\end{array}
\label{crit}
\ee

To find the corrections to scaling in the form of $\ell(\omega)$, we first note that it is
given by the bare length-scale of renormalized sites once the RG scale reaches
$\Gamma=\ln\frac{\omega_0}{\omega}$, with $\omega_0$ the bare energy
scale of the Bose-Hubbard chain. The RG flow of the effective site
and bond length is:
\be
\frac{d\ell}{d\Gamma}=\ell(f+g).
\label{lrg}
\ee
At the critical point we expect
$\ell(\omega)\sim{1/\omega}$; let us first derive the correction
to scaling at the critical point. Integrating Eq. (\ref{lrg}) using
Eq. (\ref{crit}) gives $\ln \ell=\Gamma+2\ln \Gamma/\Gamma_0 +{\cal
  O}(1/\Gamma)$, and thus we find the localization length at
criticality having a logarithmic correction:
\be
\ell(\omega)\sim \left[\ln^2\left(\omega/\omega_0\right) \right]/\omega.
\label{loctran}
\ee
Off criticality, we find by the same analysis:
\be
\ell(\omega)\sim
\left[ \left( 1- \left(\omega/\omega_0 \right)^\epsilon \right)/\epsilon \right]^2 /\omega^{1+\epsilon}.
\ee

\begin{figure}
\includegraphics[width=6.95cm]{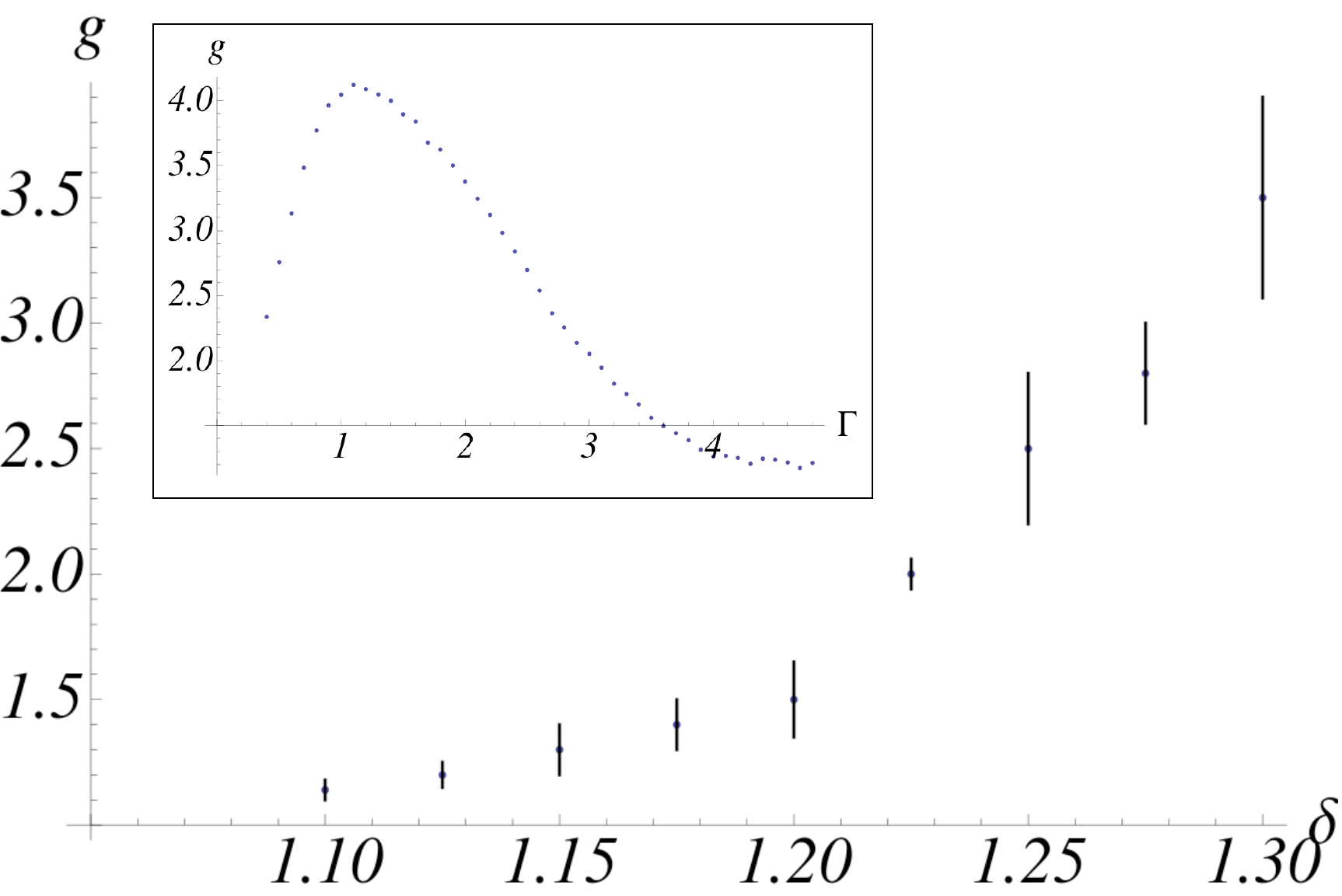}
\caption{An example of the correspondence between initial
  distributions, and the exponent $g=\alpha$ at the end of the
  flow. \textbf{Main plot:} terminal $g$ as a function of the
  parameter $\delta$ for initial Gaussian coupling distributions
  $P_U(U)\sim e^{-(U-0.3)^2/0.7^2}$, and $P_J(J)\sim
  e^{-(J-\delta)^2/0.4^2}$, truncated for $J,U<10^{-4}$. \textbf{Inset:}
example of the flow of $g$ vs. the RG flow parameter
  $\Gamma$ for the Gaussian initial conditions with
  $\delta=1.125$. The localization exponent $\alpha$ at $\delta=1.25$
  is $g$ at the large $\Gamma$ plateau. Each point is averaged over 40 disorder 
  realizations, with the chains of $5\cdot 10^6$ sites. \label{fig1}}
\end{figure}

In summary, localization properties at low frequency are determined by the parameter $g$. Its
value cannot be calculated directly in closed form from the initial disorder distribution,
but we can estimate it by following the RG
flow using the techniques of Refs.~\cite{Altman2004,Altman2007}. In Fig. \ref{fig1} we demonstrate how initial distributions
evolve into the exponent $\alpha$, which is $g$ at the end of the
flow. 

Finally, we remark that the results of this paper should also be valid in the case of the quasi one-dimensional condensates in the
presence of random potential (but in the absence of any lattice). Indeed, such condensates are expected to form puddles in the minima of the
potential, with nonzero boson hopping amplitudes  between the puddles. Then they are expected to mimic \rf{eq:ham}, and
the rest of the analysis of this paper applies.

We acknowledge support from the NSF grant DMR-0449521 and from the NIST - CU seed grant (VG), and from 
the EPSRC Grant No. EP/D050952/1 (JTC).

\bibliography{dyson}

\begin{thebibliography}{21}
\expandafter\ifx\csname natexlab\endcsname\relax\def\natexlab#1{#1}\fi
\expandafter\ifx\csname bibnamefont\endcsname\relax
  \def\bibnamefont#1{#1}\fi
\expandafter\ifx\csname bibfnamefont\endcsname\relax
  \def\bibfnamefont#1{#1}\fi
\expandafter\ifx\csname citenamefont\endcsname\relax
  \def\citenamefont#1{#1}\fi
\expandafter\ifx\csname url\endcsname\relax
  \def\url#1{\texttt{#1}}\fi
\expandafter\ifx\csname urlprefix\endcsname\relax\def\urlprefix{URL }\fi
\providecommand{\bibinfo}[2]{#2}
\providecommand{\eprint}[2][]{\url{#2}}

\bibitem[{\citenamefont{Haviland et~al.}(1989)\citenamefont{Haviland, Liu, and
  Goldman}}]{GoldmanHaviland}
\bibinfo{author}{\bibfnamefont{D.~B.} \bibnamefont{Haviland}},
  \bibinfo{author}{\bibfnamefont{Y.}~\bibnamefont{Liu}}, \bibnamefont{and}
  \bibinfo{author}{\bibfnamefont{A.~M.} \bibnamefont{Goldman}},
  \bibinfo{journal}{Phys. Rev. Lett.} \textbf{\bibinfo{volume}{62}},
  \bibinfo{pages}{2180} (\bibinfo{year}{1989}).

\bibitem[{\citenamefont{Frydman et~al.}(2002)\citenamefont{Frydman, Naaman, and
  Dynes}}]{DynesG}
\bibinfo{author}{\bibfnamefont{A.}~\bibnamefont{Frydman}},
  \bibinfo{author}{\bibfnamefont{O.}~\bibnamefont{Naaman}}, \bibnamefont{and}
  \bibinfo{author}{\bibfnamefont{R.~C.} \bibnamefont{Dynes}},
  \bibinfo{journal}{Phys. Rev. B} \textbf{\bibinfo{volume}{66}},
  \bibinfo{pages}{052509} (\bibinfo{year}{2002}).

\bibitem[{\citenamefont{Lau et~al.}(2001)\citenamefont{Lau, Markovic, Bockrath,
  Bezryadin, and Tinkham}}]{Lau}
\bibinfo{author}{\bibfnamefont{C.~N.} \bibnamefont{Lau}},
  \bibinfo{author}{\bibfnamefont{N.}~\bibnamefont{Markovic}},
  \bibinfo{author}{\bibfnamefont{M.}~\bibnamefont{Bockrath}},
  \bibinfo{author}{\bibfnamefont{A.}~\bibnamefont{Bezryadin}},
  \bibnamefont{and} \bibinfo{author}{\bibfnamefont{M.}~\bibnamefont{Tinkham}},
  \bibinfo{journal}{Phys. Rev. Lett.} \textbf{\bibinfo{volume}{87}},
  \bibinfo{pages}{217003} (\bibinfo{year}{2001}).

\bibitem[{\citenamefont{Crowell et~al.}(1997)\citenamefont{Crowell, Van~Keuls,
  and Reppy}}]{Reppy2}
\bibinfo{author}{\bibfnamefont{P.~A.} \bibnamefont{Crowell}},
  \bibinfo{author}{\bibfnamefont{F.~W.} \bibnamefont{Van~Keuls}},
  \bibnamefont{and} \bibinfo{author}{\bibfnamefont{J.~D.} \bibnamefont{Reppy}},
  \bibinfo{journal}{Phys. Rev. B} \textbf{\bibinfo{volume}{55}},
  \bibinfo{pages}{12620} (\bibinfo{year}{1997}).

\bibitem[{\citenamefont{Sanchez-Palencia
  et~al.}(2007)\citenamefont{Sanchez-Palencia, Cl\'ement, Lugan, Bouyer,
  Shlyapnikov, and Aspect}}]{Aspect2006}
\bibinfo{author}{\bibfnamefont{L.}~\bibnamefont{Sanchez-Palencia}},
  \bibinfo{author}{\bibfnamefont{D.}~\bibnamefont{Cl\'ement}},
  \bibinfo{author}{\bibfnamefont{P.}~\bibnamefont{Lugan}},
  \bibinfo{author}{\bibfnamefont{P.}~\bibnamefont{Bouyer}},
  \bibinfo{author}{\bibfnamefont{G.}~\bibnamefont{Shlyapnikov}},
  \bibnamefont{and} \bibinfo{author}{\bibfnamefont{A.}~\bibnamefont{Aspect}},
  \bibinfo{journal}{Phys. Rev. Lett.} \textbf{\bibinfo{volume}{98}},
  \bibinfo{pages}{210401} (\bibinfo{year}{2007}).

\bibitem[{\citenamefont{Bilas and Pavloff}(2006)}]{Pavloff2006}
\bibinfo{author}{\bibfnamefont{N.}~\bibnamefont{Bilas}} \bibnamefont{and}
  \bibinfo{author}{\bibfnamefont{N.}~\bibnamefont{Pavloff}},
  \bibinfo{journal}{Eur. Phys. J. D.} \textbf{\bibinfo{volume}{40}},
  \bibinfo{pages}{387} (\bibinfo{year}{2006}).

\bibitem[{\citenamefont{Lugan et~al.}(2007)\citenamefont{Lugan, Cl\'ement,
  Bouyer, Aspect, and Sanchez-Palencia}}]{Aspect2007}
\bibinfo{author}{\bibfnamefont{P.}~\bibnamefont{Lugan}},
  \bibinfo{author}{\bibfnamefont{D.}~\bibnamefont{Cl\'ement}},
  \bibinfo{author}{\bibfnamefont{P.}~\bibnamefont{Bouyer}},
  \bibinfo{author}{\bibfnamefont{A.}~\bibnamefont{Aspect}}, \bibnamefont{and}
  \bibinfo{author}{\bibfnamefont{L.}~\bibnamefont{Sanchez-Palencia}},
  \bibinfo{journal}{Phys. Rev. Lett.} \textbf{\bibinfo{volume}{99}},
  \bibinfo{pages}{180402} (\bibinfo{year}{2007}).

\bibitem[{\citenamefont{Fallani et~al.}(2007)\citenamefont{Fallani, Lye,
  Guarrera, Fort, and Inguscio}}]{Inguscio2007}
\bibinfo{author}{\bibfnamefont{L.}~\bibnamefont{Fallani}},
  \bibinfo{author}{\bibfnamefont{J.~E.} \bibnamefont{Lye}},
  \bibinfo{author}{\bibfnamefont{V.}~\bibnamefont{Guarrera}},
  \bibinfo{author}{\bibfnamefont{C.}~\bibnamefont{Fort}}, \bibnamefont{and}
  \bibinfo{author}{\bibfnamefont{M.}~\bibnamefont{Inguscio}},
  \bibinfo{journal}{Phys. Rev. Lett.} \textbf{\bibinfo{volume}{98}},
  \bibinfo{pages}{130404} (\bibinfo{year}{2007}).

\bibitem[{\citenamefont{Lye et~al.}(2007)\citenamefont{Lye, Fallani, Fort,
  Guarrera, Mudugno, Wiersma, and Inguscio}}]{Inguscio2007a}
\bibinfo{author}{\bibfnamefont{J.~E.} \bibnamefont{Lye}},
  \bibinfo{author}{\bibfnamefont{L.}~\bibnamefont{Fallani}},
  \bibinfo{author}{\bibfnamefont{C.}~\bibnamefont{Fort}},
  \bibinfo{author}{\bibfnamefont{V.}~\bibnamefont{Guarrera}},
  \bibinfo{author}{\bibfnamefont{M.}~\bibnamefont{Mudugno}},
  \bibinfo{author}{\bibfnamefont{D.~S.} \bibnamefont{Wiersma}},
  \bibnamefont{and} \bibinfo{author}{\bibfnamefont{M.}~\bibnamefont{Inguscio}},
  \bibinfo{journal}{Phys. Rev. A} \textbf{\bibinfo{volume}{75}},
  \bibinfo{pages}{061603} (\bibinfo{year}{2007}).

\bibitem[{\citenamefont{Chow et~al.}(1998)\citenamefont{Chow, Delsing, and
  Haviland}}]{Delsing1998}
\bibinfo{author}{\bibfnamefont{E.}~\bibnamefont{Chow}},
  \bibinfo{author}{\bibfnamefont{P.}~\bibnamefont{Delsing}}, \bibnamefont{and}
  \bibinfo{author}{\bibfnamefont{D.~B.} \bibnamefont{Haviland}},
  \bibinfo{journal}{Phys. Rev. Lett.} \textbf{\bibinfo{volume}{81}},
  \bibinfo{pages}{204} (\bibinfo{year}{1998}).

\bibitem[{\citenamefont{Castellanos-Beltran and Lehnert}(1997)}]{Lehnert2007}
\bibinfo{author}{\bibfnamefont{M.~A.} \bibnamefont{Castellanos-Beltran}}
  \bibnamefont{and} \bibinfo{author}{\bibfnamefont{K.~W.}
  \bibnamefont{Lehnert}} (\bibinfo{year}{1997}), \eprint{arXiv:0706.2373}.

\bibitem[{\citenamefont{Rimberg et~al.}(1997)\citenamefont{Rimberg, Ho, Kurdak,
  Clarke, Campman, and Gossard}}]{Clarke}
\bibinfo{author}{\bibfnamefont{A.~J.} \bibnamefont{Rimberg}},
  \bibinfo{author}{\bibfnamefont{T.~R.} \bibnamefont{Ho}},
  \bibinfo{author}{\bibfnamefont{C.}~\bibnamefont{Kurdak}},
  \bibinfo{author}{\bibfnamefont{J.}~\bibnamefont{Clarke}},
  \bibinfo{author}{\bibfnamefont{K.~L.} \bibnamefont{Campman}},
  \bibnamefont{and} \bibinfo{author}{\bibfnamefont{A.~C.}
  \bibnamefont{Gossard}}, \bibinfo{journal}{Phys. Rev. Lett.}
  \textbf{\bibinfo{volume}{78}}, \bibinfo{pages}{2632} (\bibinfo{year}{1997}).

\bibitem[{\citenamefont{Giamarchi and Schultz}(1988)}]{Giamarchi1988}
\bibinfo{author}{\bibfnamefont{T.}~\bibnamefont{Giamarchi}} \bibnamefont{and}
  \bibinfo{author}{\bibfnamefont{H.}~\bibnamefont{Schultz}},
  \bibinfo{journal}{Phys. Rev. B} \textbf{\bibinfo{volume}{37}},
  \bibinfo{pages}{325} (\bibinfo{year}{1988}).

\bibitem[{\citenamefont{Fisher et~al.}(1989)\citenamefont{Fisher, Weichman,
  Grinstein, and Fisher}}]{Weichman1989}
\bibinfo{author}{\bibfnamefont{M.~P.~A.} \bibnamefont{Fisher}},
  \bibinfo{author}{\bibfnamefont{P.~B.} \bibnamefont{Weichman}},
  \bibinfo{author}{\bibfnamefont{G.}~\bibnamefont{Grinstein}},
  \bibnamefont{and} \bibinfo{author}{\bibfnamefont{D.~S.}
  \bibnamefont{Fisher}}, \bibinfo{journal}{Phys. Rev. B}
  \textbf{\bibinfo{volume}{40}}, \bibinfo{pages}{546} (\bibinfo{year}{1989}).

\bibitem[{\citenamefont{Hoyos et~al.}(2007)\citenamefont{Hoyos, Kotabage, and
  Vojta}}]{Vojta2007}
\bibinfo{author}{\bibfnamefont{J.~A.} \bibnamefont{Hoyos}},
  \bibinfo{author}{\bibfnamefont{C.}~\bibnamefont{Kotabage}}, \bibnamefont{and}
  \bibinfo{author}{\bibfnamefont{T.}~\bibnamefont{Vojta}},
  \bibinfo{journal}{Phys. Rev. Lett.} \textbf{\bibinfo{volume}{99}},
  \bibinfo{pages}{230601} (\bibinfo{year}{2007}).

\bibitem[{\citenamefont{Altman et~al.}(2004)\citenamefont{Altman, Kafri,
  Polkovnikov, and Refael}}]{Altman2004}
\bibinfo{author}{\bibfnamefont{Y.}~\bibnamefont{Altman}},
  \bibinfo{author}{\bibfnamefont{Y.}~\bibnamefont{Kafri}},
  \bibinfo{author}{\bibfnamefont{A.}~\bibnamefont{Polkovnikov}},
  \bibnamefont{and} \bibinfo{author}{\bibfnamefont{G.}~\bibnamefont{Refael}},
  \bibinfo{journal}{Phys. Rev. Lett.} \textbf{\bibinfo{volume}{93}},
  \bibinfo{pages}{150402} (\bibinfo{year}{2004}).

\bibitem[{\citenamefont{Altman et~al.}(2008)\citenamefont{Altman, Kafri,
  Polkovnikov, and Refael}}]{Altman2007}
\bibinfo{author}{\bibfnamefont{Y.}~\bibnamefont{Altman}},
  \bibinfo{author}{\bibfnamefont{Y.}~\bibnamefont{Kafri}},
  \bibinfo{author}{\bibfnamefont{A.}~\bibnamefont{Polkovnikov}},
  \bibnamefont{and} \bibinfo{author}{\bibfnamefont{G.}~\bibnamefont{Refael}},
  \bibinfo{journal}{Phys. Rev. Lett.} \textbf{\bibinfo{volume}{100}},
  \bibinfo{pages}{170402} (\bibinfo{year}{2008}).

\bibitem[{\citenamefont{Ishii}(1973)}]{Ishii1973}
\bibinfo{author}{\bibfnamefont{K.}~\bibnamefont{Ishii}},
  \bibinfo{journal}{Prog. Theor. Phys. Suppl.} \textbf{\bibinfo{volume}{53}},
  \bibinfo{pages}{77} (\bibinfo{year}{1973}).

\bibitem[{\citenamefont{Ziman}(1982)}]{Ziman1982}
\bibinfo{author}{\bibfnamefont{T.}~\bibnamefont{Ziman}},
  \bibinfo{journal}{Phys. Rev. Lett.} \textbf{\bibinfo{volume}{49}},
  \bibinfo{pages}{337} (\bibinfo{year}{1982}).

\bibitem[{\citenamefont{John et~al.}(1982)\citenamefont{John, Sompolinsky, and
  Stephen}}]{John1982}
\bibinfo{author}{\bibfnamefont{S.}~\bibnamefont{John}},
  \bibinfo{author}{\bibfnamefont{H.}~\bibnamefont{Sompolinsky}},
  \bibnamefont{and} \bibinfo{author}{\bibfnamefont{M.~J.}
  \bibnamefont{Stephen}}, \bibinfo{journal}{Phys. Rev. B}
  \textbf{\bibinfo{volume}{27}}, \bibinfo{pages}{5592} (\bibinfo{year}{1982}).

\bibitem[{\citenamefont{Gurarie and Chalker}(2003)}]{Gurarie2003a}
\bibinfo{author}{\bibfnamefont{V.}~\bibnamefont{Gurarie}} \bibnamefont{and}
  \bibinfo{author}{\bibfnamefont{J.~T.} \bibnamefont{Chalker}},
  \bibinfo{journal}{Phys. Rev. B} \textbf{\bibinfo{volume}{68}},
  \bibinfo{pages}{134207} (\bibinfo{year}{2003}).

\end{thebibliography}

\end{document}